%% file: main.tex
\documentclass[conference]{IEEEtran}
\usepackage{cite}
\usepackage{amsmath,amssymb,amsfonts}
\usepackage{algorithmic}
\usepackage[bookmarks=false]{hyperref}
\usepackage{listings}
\usepackage{graphicx}
\usepackage{xcolor}
\usepackage{tabularx}
\usepackage{tabulary}
\usepackage{array}
\usepackage{booktabs}
\usepackage{pifont}             
\usepackage[shortlabels]{enumitem}			
\usepackage{makecell} 

\usepackage{xcolor} 
\definecolor{light-gray}{gray}{0.95}

\begin{document}
\bstctlcite{IEEEexample:BSTcontrol} 

\title{A Mapping Approach to Convert MTPs into a Capability and Skill Ontology}

\author{
\IEEEauthorblockN{
    Aljosha Köcher\IEEEauthorrefmark{1},
    Lasse Beers\IEEEauthorrefmark{1},
    Alexander Fay\IEEEauthorrefmark{1},
}
\IEEEauthorblockA{
\IEEEauthorrefmark{1}Institute of Automation\\
Helmut Schmidt University, Hamburg, Germany\\
Email: aljosha.koecher@hsu-hh.de, lasse.beers@hsu-hh.de, alexander.fay@hsu-hh.de\\}
}

\maketitle

\begin{abstract}
Being able to quickly integrate new equipment
and functions into an existing plant is a major goal for both discrete and process manufacturing. 
But currently, these two industry domains use different approaches to achieve this goal. 
While the Module Type Package (MTP) is getting more and more adapted in practical applications of process manufacturing, so-called skill-based manufacturing approaches are favored in the context of discrete manufacturing. 
The two approaches are incompatible because their models feature different contents and they use different technologies. 
This contribution provides a comparison of the MTP with a skill-based approach as well as an automated mapping that can be used to transfer the contents of an MTP into a skill ontology.
Through this mapping, an MTP can be semantically lifted in order to apply functions like querying or reasoning.
Furthermore, machines that were previously described using two incompatible models can now be used in one production process.
\end{abstract}


\section{Introduction}
\label{sec:Introduction}
\input{01_introduction}

\section{Related Work}
\label{sec:relatedWork}
\input{02_relatedWork}

\section{Modeling machine functions}
\label{sec:fundamentals}
\input{03_fundamentals}

\section{Mapping Approach}
\label{sec:mappingApproach}
\input{04_mappingApproach}

\section{Evaluation}
\label{sec:evaluation}
\input{05_evaluation}

\section{Summary \& Outlook}
\label{sec:summary}
\input{06_summaryOutlook}

\bibliographystyle{IEEEtran}
\bibliography{references}

\end{document}

%% file: 01_introduction.tex
Companies in both discrete and process manufacturing face a variety of disruptive challenges. Resource scarcity as well as increasing energy demands call for sustainable processes. At the same time, global markets generate a need for shorter product life cycles which in turn require a decrease in time to market for new products. These challenges cannot be met with inflexible production infrastructures, which are typically very limited in their adaptability. \cite{Buc_Futuremanufacturingapproachesin_2010}
Thus, there is a need for robust production systems that can cope with changes in demand and introduction of new products and processes.\cite{_BeschreibungundMessungder_2017}
Modular plants and simple methods to integrate new equipment are strived for. Such methods are grouped under the term \emph{Plug and Produce} \cite{AAM+_AgileAssemblySystemby_2000}. Plug and Produce requires a machine-readable description of modules and their functions, because only with such a description is it possible to easily adapt plants by adding or removing modules or individual functions \cite{CFL_Dynamicskillallocationmethodology_2013}.

While in process manufacturing, the \emph{Module Type Package} (MTP) is standardized by working groups of the organisations VDI, VDE, ZVEI and NAMUR, in discrete manufacturing, such machine-readable descriptions are being researched by the term \emph{capability- and skill-based manufacturing} \cite{Jar_Capabilitybasedadaptationofproduction_2012, WBS+_AnOntologybasedMetamodelfor_2020, KHV+_AFormalCapabilityand_2020}. 
In this context, a \emph{capability} describes a function that a machine is able to perform \cite{Jar_Capabilitybasedadaptationofproduction_2012}. Skills are implementations of such functions with an invocation interface (e.g. via OPC UA).
It is shown in \cite{KHV+_AFormalCapabilityand_2020} that the requirements against capabilities and skills necessitate semantic models, i.e., ontologies.

In this contribution, the term \emph{ontology} is used for highly formal ontologies defined in the \emph{Web Ontology Language} (OWL) \cite{W3C_OWL2WebOntology_2012}.
Ontologies include a set of concepts with a specification of their meaning along with definitions of how concepts are related \cite{Usc_Knowledgelevelmodelling:concepts_1998}.
Ontologies can be divided into a terminological box (T-Box) and an assertional box (A-Box). While a T-Box entails knowledge about a whole problem domain, an A-Box contains knowledge of one specific problem \cite{BCM+_TheDescriptionLogicHandbook_2007}.
OWL is an important technology in the Semantic Web, which includes other powerful functionalities such as querying models using \emph{SPARQL} or inferring new knowledge using reasoners.

While many approaches of capability- and skill-based manufacturing use ontologies, the MTP is based on AutomationML (AML), a data format based on XML. 
In addition to the modeling technology, the contents of capability/skill models and MTPs differ greatly.
Thus, MTPs and capability/skill models are currently not interoperable. While there are existing works to extend the use of MTPs from purely continuous processes to subsequent logistic operations \cite{BFG+_DesignprinzipienfurdenModulund_2021}, this contribution takes a slightly different approach:
The research objective of this contribution is to present an automatic transformation from MTPs into an ontological capability and skill model. 
We aim to achieve the following goals:
\begin{enumerate}
    \item A comparison of MTPs with an existing semantic capability/skill model.
    \item A \emph{semantic lifting} of MTPs which allows to apply functions of the Semantic Web -- e.g., querying or reasoning.
    \item Skill-based control of process modules (originally described with MTPs).
\end{enumerate}

The remainder of this paper begins with an overview of related works. In Section~\ref{sec:fundamentals}, both the MTP and a capability/skill model are introduced and a comparison is made. 
The actual mapping approach is presented in Section~\ref{sec:mappingApproach}. Section~\ref{sec:evaluation} contains an evaluation before Section~\ref{sec:summary} concludes this paper with a summary and outlook.

%% file: 02_relatedWork.tex
Related approaches are covered in two different subsections. While the first one is concerned with the research area of capabilities and skills, the second one covers automatic transformations of XML files into ontologies.

\subsection{Capabilities and Skills}
Research in capability and skill based engineering represents a rather new field with first approaches using the term \emph{skill} in the context of machine functions dating back to 2005 \cite{BCO_Amultiagentbasedcontrol_2005}. Since then, this topic has been becoming more and more relevant \cite{FKM+_CapabilitiesAndSkillsIn_2022}.

An early approach of a formal description of skills is shown in \cite{CaBa_AMultiagentControlSystem_2007}, where an ontology is used to describe both simple as well as complex skills. The focus of this work is on an implementation using a Multi-Agent-System and the actual skill ontology is presented only briefly. 
Approaches to capability description have favored formal methods since the emergence of this field of research. In \cite{WBS+_AnOntologybasedMetamodelfor_2020}, a rather generic higher-level ontology architecture is presented that can be used to model capabilities.

A group of researchers at Tampere University, Finland, has published several contributions on formal capability descriptions.
Ref. \cite{Jar_Capabilitybasedadaptationofproduction_2012} presents an automated adaptation method that is based on an ontological capability model. 
Capabilities are described by their name and parameters. Based on \cite{Jar_Capabilitybasedadaptationofproduction_2012}, sophisticated matchmaking rules that can be used to automatically generate possible resource combinations for a given production requirement were later presented \cite{JSO+_CapabilityMatchmakingProcedureto_2017}.

Most approaches that focus on the aspect of executable skills use less formal models than ontologies. Functions blocks as defined in IEC 61499 are used by \cite{FeLo_Configurationmodelforevolvable_2012} to model and execute skills. 
The authors of \cite{FeLo_Configurationmodelforevolvable_2012} later extend their approach by using an AutomationML model to represent skills in a machine-interpretable way \cite{DFL+_AnAutomationMLmodelfor_2017}.

In \cite{KHV+_AFormalCapabilityand_2020}, the benefits of joining both capability and skill aspects in one coherent semantic model are discussed and an ontology is presented that consists of individual so-called \emph{Ontology Design Patterns} (ODPs), which are based on standards. This model combines the benefits of previously separated approaches to capability and skill modeling.

Despite the large number of contributions dealing with models of capabilities and skills, there are very few methods and tools to create these models efficiently and with little effort.
In \cite{KHC+_AutomatingtheDevelopmentof_2020}, a method to largely automate creation of a capability/skill model is presented. With this method, users are assisted in creating capabilities and skills are implemented using the Java framework \emph{SkillUp}\footnotemark{}. All tasks required to create an ontological model are fully automated.
In \cite{KJF_AMethodtoAutomatically_2021}, the approach of \cite{KHC+_AutomatingtheDevelopmentof_2020} was extended to PLCs. Through an automated mapping based on the RDF Mapping Language (RML), a skill ontology is created for existing PLC programs.

\footnotetext{SkillUp is maintained as an open-source project at https://github.com/aljoshakoecher/skill-up}

\subsection{Automated Mapping to Ontologies}
Because there are so few approaches to automatically create capability and skill models, we present more general contributions that automatically map data from XML sources into an ontology.

An early approach to transform an XML file into an ontology is presented in \cite{rodrigues2006mapping} in which an existing ontology is filled with content through a mapping. The approach assumes that an XML schema of the source file exists. 
One of the first approaches with a more specific CAEX transformation is shown in \cite{Runde.2009}. The authors transfer CAEX system data into an OWL ontology. The data of the role and interface class library are represented on a terminological level. Further information about the element type of the CAEX schema is attached to the role or interface library as meta-information. 

The focus of \cite{Abele.29.07.201331.07.2013} is on using SPARQL for automated validation of CAEX plant data. With an example of a CAEX plant model, the authors show how inconsistencies can be detected and corrected by validating existing plant data. A basic ontology is presented, which contains essential CAEX terms and is able to capture a basic ontological plant representation. 
The authors of \cite{StefanZanderandYingbingHua.2017} focus primarily on reasoning. The approach is concerned with supplementing the technical specifications of hardware components by using ontological classifications and arguments. In \cite{Hua.10.09.201913.09.2019}, the authors criticize that previous approaches only cover simple knowledge structures and present an approach to transfer complex OWL classes into AML. A reverse transformation is also discussed.

In summary, previous contributions developed transformations because of a lack of semantics in CAEX. The approaches of \cite{Abele.29.07.201331.07.2013} and \cite{Runde.2009} attempt to transfer the CAEX schema as accurately as possible. 
To our knowledge, no approach exists that transforms an MTP into an ontology. The approach taken in this work aims to convert the service description of an MTP into an existing capability/skill ontology with a given T-Box. Thus, mapping rules are needed that transfer CAEX model elements to a model corresponding to this T-Box.

%% file: 03_fundamentals.tex
\subsection{Module Type Package}
\label{subsec:mtp}
The MTP is a promising way to achieve interoperability of modules in process manufacturing as it provides a standardized and manufacturer-independent description of modules, also called \emph{Process Equipment Assemblies} (PEA) \cite{Klose.2019}. The MTP is used to describe functionalities of process modules as so-called \emph{services} in a machine-readable way to efficiently integrate modules into a so-called \emph{Process Orchestration Layer} (POL). 

Currently, the MTP includes a description of a PEA's functionalities, all communication channels and variables of the module, and all necessary information for automated generation of a \emph{human-machine interface} (HMI) as well as an alarm management concept for modular production.
Further extensions are currently being developed. 
The technical exchange format AML is used to represent MTPs in a machine-readable way. \cite{Ladiges.2018}

In the following, we will mainly focus on the structural design of the MTP, as this is crucial for our mapping approach.

AutomationML defines so-called \emph{InstanceHierarchies} (IHs) to capture objects of a model. For an MTP, different IHs are created to organize different aspects.
To describe a PEA's functionalities, the IH \emph{Services} is provided that consists of several InternalElements of predefined classes.
While the class \emph{Service} is used for module services, the class \emph{ServiceProcedure} is used to describe various operating modes that define the service in more detail. 
The \emph{ConfigurationParameter} and \emph{ProcedureParameter} classes are used to parameterize procedures and services. 
Furthermore, the classes \emph{ReportValue}, \emph{ProcessValueIn} and \emph{ProcessValueOut} can be created per procedure. While the \emph{ReportValue} class is used to document process values, the classes ProcessValueIn and ProcessValueOut are mainly used for module-to-module communication. \cite{Hoernicke.2021}

A separate IH contains object and communication information in a so-called \emph{CommunicationSet}.
This IH contains the \emph{InstanceList}, where all actuators and sensors of a module are listed. Additional elements which control and parameterize procedures and services via their attributes are also part of the InstanceList.
The \emph{SourceList} describes the communication aspect of an MTP. It defines how the communication with a PEA can be accomplished by describing a binding of all MTP attributes to an OPC UA server's address space.
To link two or more elements of an MTP, a so-called \emph{LinkedObject concept} is employed by using the same \emph{RefID} attribute for these elements. The focus of this work is on transferring service aspects and equipment (i.e., sensors/actuators) of an MTP, therefore other elements like the HMI are not considered for this approach.

\subsection{Capability and Skill Ontology}
\label{subsec:skill-model}
\begin{figure}[h]
    \centering
    \includegraphics[width=0.75\linewidth]{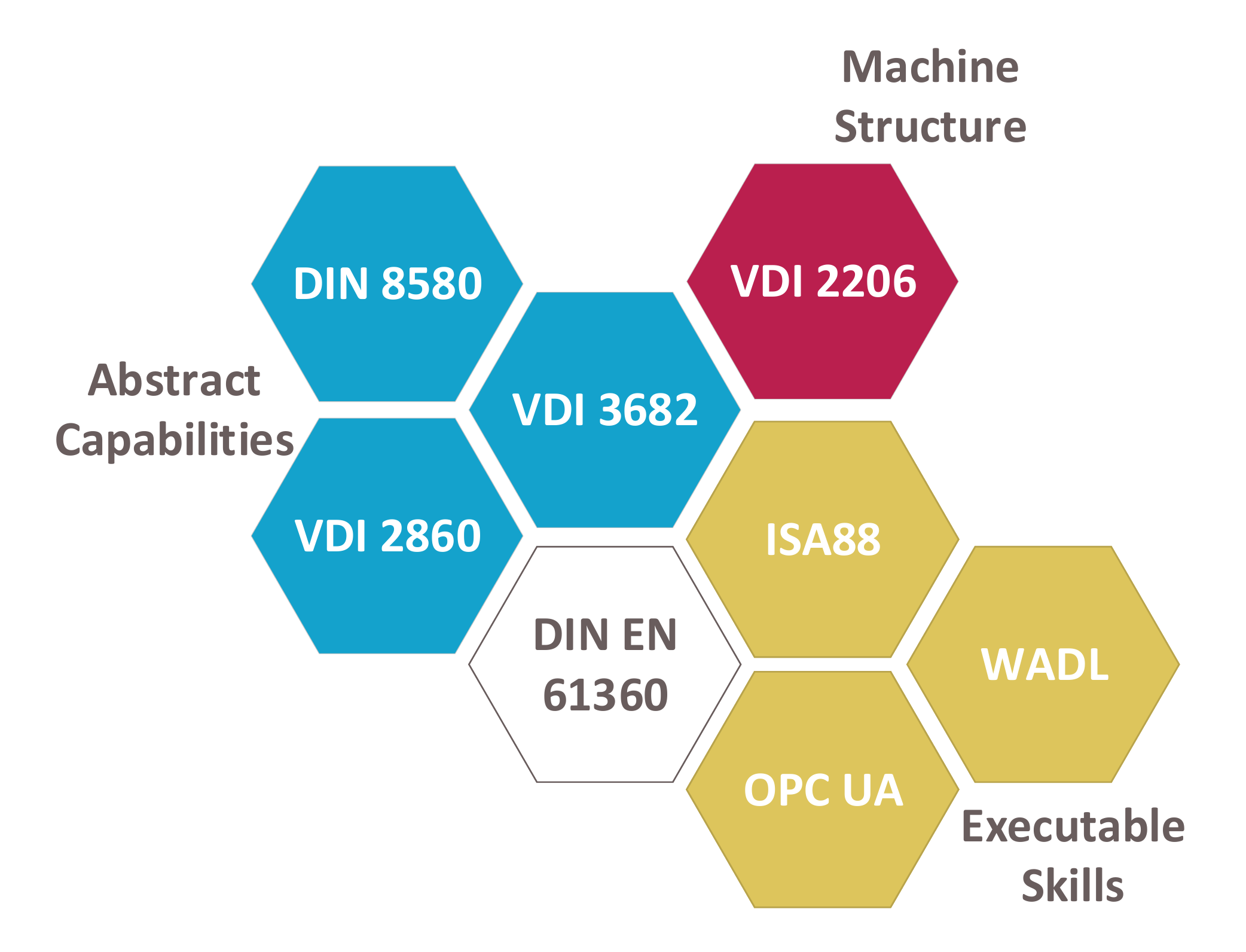}
    \caption{ODPs of the capability/skill model and the sub models \emph{machine structure} (red), \emph{ capabilities} (blue), \emph{skills} (yellow) and \emph{properties} (white).}
    \label{fig:ODPs}
\end{figure}

In our previous works, we created and used an ontology to describe modules, their capabilities and skills in a formal way \cite{KHV+_AFormalCapabilityand_2020, KHC+_AutomatingtheDevelopmentof_2020, KJF_AMethodtoAutomatically_2021}. This ontology consists of multiple distinct ontologies, so called \emph{Ontology Design Patterns} (ODPs) that are all based on industry standards\footnotemark. 
\footnotetext{https://github.com/hsu-aut/Industrial-Standard-Ontology-Design-Patterns}
Figure \ref{fig:ODPs} depicts the ODPs used for our capability and skill model. The colors define four submodels -- namely \emph{machine structure} (red), \emph{capabilities} (blue), \emph{skills} (yellow) and \emph{properties} (white)

In order to get one coherent semantic model of machines, their capabilities and skills, a so-called alignment ontology is used which combines all ODPs into one model by means of e.g. equivalent classes, subclassing or new object properties. 
The main rationale behind the model can be summed up as follows: Technical resources (i.e., systems or modules) may have capabilities. These capabilities can be connected to executable skills which can be implemented either using OPC UA or as a RESTful webservice. 
A state machine conforming with ISA 88 is used to describe the current state of a skill execution as well as possible transitions from the current state. 
Transitions are connected to SkillMethods, which can be implemented either as OPC UA methods or web services according to WADL, thus creating a binding between a skill's behavior description and a callable interface. 
Please note that details of the alignment ontology have been presented in \cite{KHV+_AFormalCapabilityand_2020} and the ontology is available online\footnote{https://github.com/aljoshakoecher/machine-skill-model}. 

\smallskip
In this section, we focus on extensions to our model that were necessary in order to model services of an MTP as skills. MTP services can currently not be invoked by calling an OPC UA method. Instead, a specific OPC UA integer variable is used to trigger transitions. Another integer variable is used to represent the current state. In order to model skills that are triggered by OPC UA variables instead of methods, we added an OWL class \verb|Cap:OpcUaVariableSkill| to our ontology. To model the trigger variable itself, the subclass \verb|Cap:SkillCommand| was added to the existing class \verb|Cap:SkillParameter|. Individuals of \verb|Cap:CurrentStateOutput| can be used for OPC UA variables that reflect the current state of a skill.

Simply adding classes for trigger and feedback variables is not enough. A machine-interpretable model has to contain information about the values such a trigger variable can take in order to invoke a skill transition. And the values which can be expected if a skill changes to a certain state have to be explicitly modelled as well. This is achieved using a property ODP according to IEC 61360 and the concepts defined in \cite{HSF+_Semanticmodelingforcollaboration_2017}. 
This ODP allows to express data property statements with different expression goals and interpretation logic. While the \verb|Expression Goal| can be one of the three \emph{Requirement}, \emph{Assurance} and \emph{Actual Value}, \verb|Interpretation Logic| is used to express relations (e.g. less than, equal to, greater than) with regard to a given value. 

\begin{figure*}[h]
    \centering
    \includegraphics[width=0.95\textwidth]{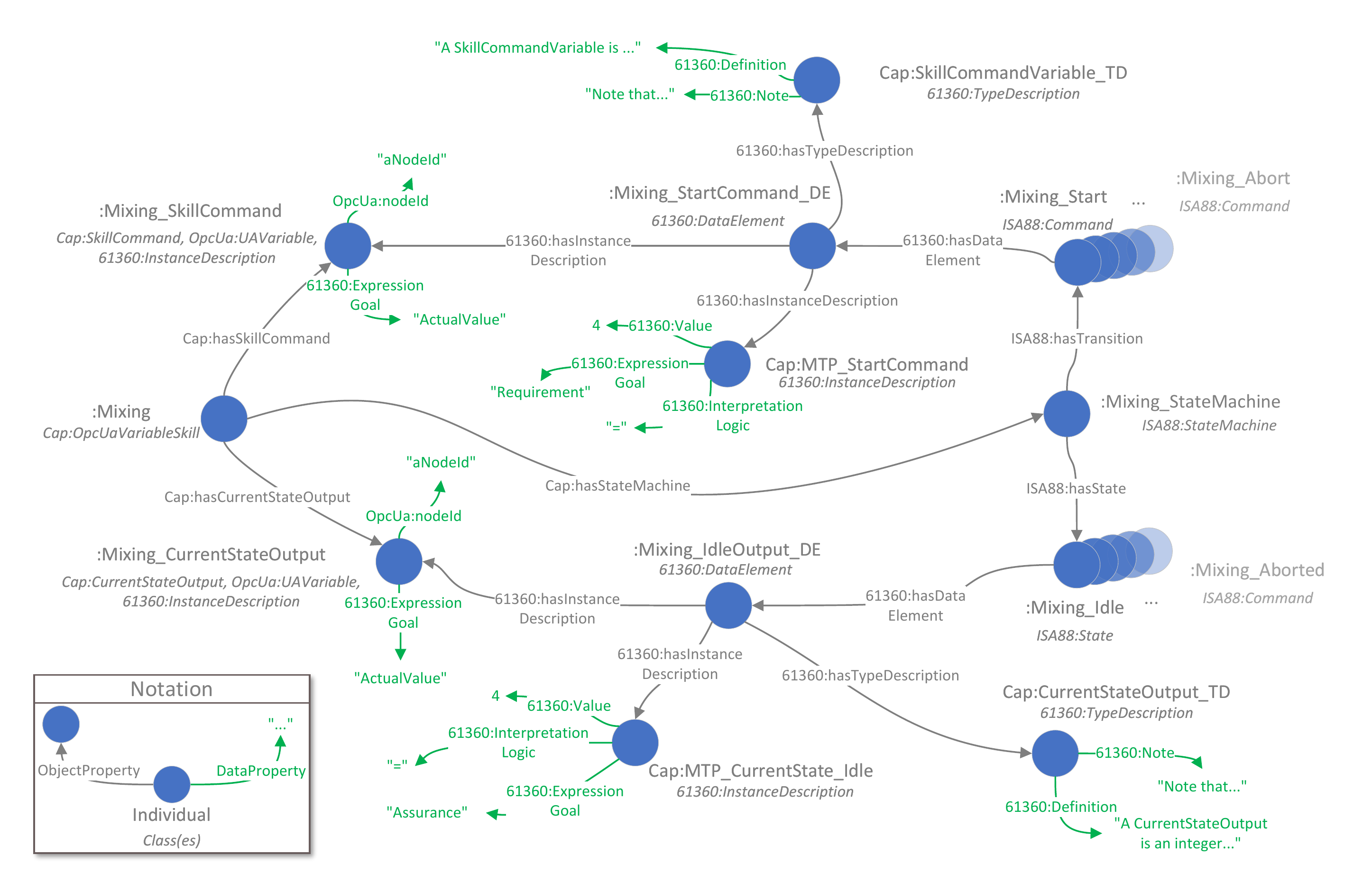}
    \caption{An excerpt of the ontology showing a skill with its parameters and their relation to the state machine. Prefixes of standards (e.g. \texttt{ISA88}) and the prefix \texttt{Cap} denote general elements of the capability/skill model while the empty prefix denotes elements of an exemplary mixing skill. 
    Some classes and properties are omitted to maintain readability.}
    \label{fig:ExampleMtpProperties}
\end{figure*}

Figure \ref{fig:ExampleMtpProperties} shows how we make use of data properties to connect a skill command parameter with transitions and a state feedback variable with states of a state machine. In this exemplary excerpt of our capability/skill model, a skill \texttt{Mixing} is depicted together with two variables: A command variable \texttt{:Mixing\_SkillCommand} and an output \texttt{:Mixing\_CurrentStateOutput}. Furthermore, the skill is connected with a state machine (\texttt{:Mixing\_StateMachine}, center right) which consists of states and transitions. All transitions and states have data elements to express commands and state feedback, respectively. All command data elements of all skills share one single type description (\texttt{Cap:SkillCommandVariable\_TD} which contains general information and a note on how to use command variables. The same holds for all data elements that represent feedback about the current state (e.g. \texttt{:Mixing\_IdleOutput\_DE}). Such data elements are connected with a single type description (\texttt{Cap:CurrentStateOutput\_TD}).

Command parameters and state outputs are modelled as both properties (\verb|DIN61360:InstanceDescription|) as well as specific OPC UA variables. 
Every \verb|OpcUaVariableSkill| has only one command parameter and one state output and these variables are connected with all data elements of transitions and states, respectively. They can be written or read through the given OPC UA variable attributes to fire a transition or get the current state's integer number, respectively.  

In addition to these instance descriptions of the OPC UA variables, other instance descriptions are used to indicate required or assured values to fire transitions or interpret state feedback, respectively. In the example of Figure \ref{fig:ExampleMtpProperties}, the data element \texttt{:Mixing\_StartCommand\_DE} is connected with \texttt{Cap:MTP\_StartCommand}. This individual can be interpreted as \emph{"In order to invoke \emph{Start}, the value of this data element is required to be equal to 4"}.


This modeling approach ensures that all information that is necessary to interact with a skill's state machine is contained in the ontology. Starting with a given value for a state feedback variable (i.e. an MTP's \emph{StateCur}), the model can be queried to find the associated state. Subsequently, possible transitions together with their required skill command values can be retrieved and invoked by setting the command variable to one of these values, depending on the desired transition.

\subsection{Comparison}
While it is obvious that both MTPs and skill models address similar goals, there is a variety of differences between these two approaches. In this subsection, a short comparison of MTPs and our capability and skill model is given.

Both approaches focus on a description of production modules and their functions and aim for an efficient integration of modules into production lines. For this purpose, automatically executable functions are encapsulated and described in a uniform way. 
While MTPs are described and stored in an AML file, our capability/skill model is an ontology modeled using OWL and stored in a graph database. 

The description of module components is done in comparable ways in both approaches. In our ontology, an ODP of VDI 2206 serves to describe structural aspects of a machine. Components are divided into actuators and sensors. An MTP contains a description of all control components that belong to a PEA in a more detailed way. The description of actuators and sensors in an MTP is done by the AML classes \emph{ActiveElements} and \emph{IndicatorElement}.

In both models, the possibility of relating two elements is important. However, because of the technologies used, this is achieved through completely different mechanisms. In an MTP, there are two ways to establish a relationship, either by using hierarchies or by using RefIDs.
In an ontology, relationships between individuals are explicitly modeled using object properties.

In our ontology, capabilities are represented by three ODPs (see Section \ref{subsec:skill-model}) allowing processes to be modeled and further subdivided into manufacturing or handling operations. 
While there are efforts to use MTPs in intralogistics (see \cite{BFG+_DesignprinzipienfurdenModulund_2021}), MTPs are targeted for process manufacturing. A classification of operations is not contained in the model. The functionalities provided by a PEA are called services and can be compared to capabilities.

In our capability model, the executable aspect of a capability is represented as a skill. 
A skill can be executed automatically thanks to a description of its invocation interface. Every capability that is also executable is connected to a skill via the object property \texttt{Cap:isExecutableVia}. A module is connected to its skills by \texttt{Cap:providesSkill}. 
In an MTP, the specification of a service is done with an internal element of the system unit class \emph{ServiceProcedure}. Every executable service must have at least one procedure, therefore a procedure can be compared to a skill.

In MTPs, there are two ways to parameterize a procedure. Configuration parameters are used to make basic settings for a service before commissioning. They always belong to exactly one service and may parameterize procedures of this service. Procedure parameters are recipe-relevant. They are assigned to one or more procedures and are written every time before a procedure is started. 
Our skill ontology offers the possibility of parameterization through individuals of the OWL class \verb|Cap:SkillParameter| but there is no parameter distinction as in MTPs.

In MTPs, the execution of a procedure is realized with a command attribute. This attribute is implemented as an OPC UA double-word variable, with each bit representing a command. Important to note is that the AutomationML model of an MTP doesn't contain the values that need to be set and read in order to trigger transitions and get the current state, respectively.
For a detailed explanation of this encoding, please refer to \cite{_Automationengineeringofmodular_2019_2658-4}. 
The \emph{CommandEn} attribute can be used to see in which state which commands are allowed. The current state can be seen by the \emph{StateCur} attribute. Each of these attributes is part of an internal element of the \emph{ServiceControl} class.

Our skill model provides an abstraction layer of various technical implementations for executing skills. Currently, OPC UA methods, OPC UA trigger variables and RESTful web services can be used. RESTful web services may not seem useful for manufacturing processes but they allow to wrap typical IT functions as skills and use them in combination with manufacturing processes.
At this time, MTPs can only be invoked using OPC UA trigger variables.

In our capability and skill ontology, the state machine of ISA88 is represented by a corresponding ODP. 
The MTP makes use of similar, but slightly different state machine. 
An important distinction between the two approaches is that the state machine of an MTP is not contained in its AML model. Our capability model, however, requires an individual \texttt{ISA88:StateMachine} with all its states and transitions for each skill.
MTPs allow control of PEAs from different sources. It is possible that a command can be requested by an operator, the POL or an internal service of the PEA. PEA control is further divided into Automatic-Internal and Automatic-External accesses.
Skills of our capability and skill model can be accessed externally -- currently without any kind of prioritization option between different types of access.
An overview of the discussed comparison between our skill ontology and MTPs can be found in Table \ref{tab:comparisonMtpSkillModel}.

\begin{table*}[h]
    \centering
    \caption{Comparison between model aspects of the MTP and our ontological capability and skill model}
    \renewcommand*{\arraystretch}{1.2}
    \begin{tabularx}{\textwidth}{X c c}
        \normalsize{Model Aspect} & \normalsize{MTP} & \normalsize{Skill Model} \\
        \hline
        Data Format & \normalfont{AML} & \normalfont{OWL} \\
        Data Storage & \normalfont{File} & \normalfont{Graph database} \\
        Service Access & \normalfont{External, Internal, Through Operator} & \normalfont{External (no differentiation)} \\
        State Machine & \normalfont{Not contained in the model} & \normalfont{Explicitly modelled} \\
        Establishing Relationships & \normalfont{RefID or Hierarchy} & \normalfont{Object Properties}\\
        Capability  & \normalfont{Service} & \normalfont{VDI3682:Process with In- and Output} \\
        Parameterisation of Capabilities & \normalfont{Configuration- or Procedure-Parameter} & \normalfont{Cap:SkillParameter} \\
        Skill & \normalfont{Procedure} & \normalfont{Cap:Skill} \\
        Command / Trigger Variables & \normalfont{Command Attribute} & \normalfont{Cap:SkillCommand}  \\
        Result Variables & \normalfont{ProcessValueOut} & \normalfont{Cap:SkillOutput} \\
    \end{tabularx}
    \label{tab:comparisonMtpSkillModel}
\end{table*}

%% file: 04_mappingApproach.tex
The \emph{RDF mapping language} (RML) is used to define a mapping between MTPs and the capability/skill model.
RML is a generic mapping language which was created to define mapping rules from heterogeneous data structures (XML, JSON, CSV) to RDF models in Turtle syntax \cite{Dimou.2014}.
An exemplary RML mapping rule is shown in Listing~\ref{lst:mapping}. After the definition of a data source, an iterator is used to loop through all elements matching a certain XPath expression (lines 4-6). 
Every element is then used to create an RDF triple, which consists of a subject (lines 8-10), predicate (line 13) and an object (lines 15-16).

In our mapping, Internal Elements (IEs) of an MTP AutomationML model and their attributes are transferred into individuals of certain classes in the ontology. The values of attributes of IEs are represented by data properties. Object properties are used to create relations between individuals.


Each MTP service is transformed into an individual of the OWL class \verb|Cap:Capability| -- and bound to the corresponding module via \verb|Cap:hasCapabilty|.
The equivalent of an MTP procedure in the capability model is a skill. Individuals of the OWL class \verb|Cap:OpcUaSkill| and \verb|VDI3682:Process| are connected with the Object Property \verb|Cap:isExecutableViaOpcUaSkill|. 

As already described in section \ref{subsec:mtp}, each service refers to an IE of the \emph{SystemUnitClass} (SUC) \emph{ServiceControl} by means of its \emph{RefID}. This \emph{ServiceControl} element has some attributes that must be transferred to the skill model (e.g., \emph{CommandExt} and \emph{StateCur}). 
The control attribute \emph{CommandExt} is transformed into an individual of the OWL class \verb|Cap:SkillCommand| and the attribute \emph{StateCur} is transformed into an individual of the OWL class \verb|Cap:CurrentStateOutput|. Both individuals are connected to the respective skill.

The values of \emph{ProcedureParameter} and \emph{ConfigurationParameter} are specified in the attributes of the corresponding IE of the type \emph{OperationElement}. Among others, the following attributes can be found there: \emph{VExt}, \emph{VMin}, \emph{VMax} and \emph{VUnit}. These are transferred into individuals of the class \verb|Cap:SkillParameter| and, in the case of the \emph{ProcedureParameter}, linked to the associated skill by the object property \verb|Cap:hasSkillParameter|. 
In the case of a configuration parameter, the individual must be linked to all skills that are subordinate to the service. 
The value of the \emph{ProcessValueOut} element is represented by the \emph{V} attribute, which is assigned to an \emph{IndicatorElement} within the InstanceList. Accordingly, an individual of \verb|Cap:SkillOutput| is created for each of these attributes. \emph{ProcessValueIn} and \emph{ReportValue} attributes can be transformed in the same way.
As mentioned in section \ref{subsec:mtp}, there must be a communication interface for the MTP. The MTP uses an IE of the SUC \emph{OPCUAServer}, which is subordinated to the \emph{SourceList} and has the attribute \emph{Endpoint}. 
With our mapping approach, this IE is transformed into one individual of the OWL class \verb|OPCUA:UaServer| with a \verb|OPCUA:UaNodeSet| instance holding all OPC UA nodes of one MTP in the ontology. In addition, for the individual representing an OPC UA server, the data property \verb|OPCUA:endpointUrl| is used to specify the server's address. 

\begin{table}[h]
    \centering
    \caption{Overview of the transferred individuals}
    \renewcommand*{\arraystretch}{1.2}
    \begin{tabular}{l r}
        \normalsize{MTP} & \normalsize{Skill Model} \\
        \hline
        \normalfont{IE ModuleTypePackage} & \normalfont{VDI2206:Module} \\
        \normalfont{IE Service} & \normalfont{VDI3682:Process / Cap:Capability} \\
        \normalfont{IE ServiceProcedure} & \normalfont{Cap:OpcUaSkill} \\
        \normalfont{IE IndicatorElement} & \normalfont{VDI2206:Sensor} \\
        \normalfont{IE ActiveElement} & \normalfont{VDi2206:Actuator} \\
        \normalfont{IE OPCUAServer} & \normalfont{OpcUa:UaServer \& NodeSet} \\
        \normalfont{Attribute VExt} & \normalfont{Cap:SkillParameter} \\
        \normalfont{Attribute VMax} & \normalfont{Cap:SkillParameter}  \\
        \normalfont{Attribute VMin} & \normalfont{Cap:SkillParameter} \\
        \normalfont{Attribute VUnit} & \normalfont{Cap:SkillParameter} \\
        \normalfont{Attribute ProcedureExt} & \normalfont{Cap:SkillParameter} \\
        \normalfont{Attribute CommandExt} & \normalfont{Cap:SkillCommand} \\
        \normalfont{Attribute StateCur} & \normalfont{Cap:CurrentStateOutput} \\
        \normalfont{Attribute ProcedureCur} & \normalfont{Cap:SkillOutput} \\
        \normalfont{Attribute ProcedureReq} & \normalfont{Cap:SkillOutput} \\
        \normalfont{Attribute V} & \normalfont{Cap:SkillOutput} \\
    \end{tabular}
    \label{tab:transferredIndividuals}
\end{table}

The AML InterfaceClass \emph{OPCUAItem}, which is used to describe OPC UA nodes in an MTP, contains the attributes \emph{Access}, \emph{Namespace} and \emph{Identifier}. These attributes are transformed into data properties and assigned to the corresponding individuals of the ontology. 
Table \ref{tab:transferredIndividuals} shows an overview of all transformed components of an MTP. 

While the ontological capability/skill model requires a separate state machine for each skill according to ISA88, the state machine of the MTP is not contained in an MTP's AML file and therefore needs to be added.
Our mapping approach automatically takes care of that for every skill. For every MTP procedure that is mapped to a skill, all states and transitions as well as the relations between them are created with a unique identifier. 

One rule to map all MTP procedures to skills and attach the necessary command variables is presented in Listing~\ref{lst:mapping} and now explained in more detail.
The \verb|rml:iterator| (line 6) refers to all procedures of a given MTP by first tracing the path to all Internal Elements of the SUC \emph{Service} before searching for all child elements of the SUC \emph{ServiceProcedure}. The \verb|rr:subjectMap| then defines an IRI for every individual and assigns the OWL class \verb|Cap:OpcUaVariableSkill| (lines 8-10). 
Creating a \verb|Cap:SkillCommand| individual for every skill and relating it to its corresponding skill is done with a \verb|rr:predicateObjectMap|.
The inner expression looks for a value, which is further processed by the expression in line 16. The command \verb|current()| refers to the current element of the \verb|rml:iterator|. The inner expression searches for the parent element of the procedure - the service - and refers to the value of its RefID attribute. Exactly this RefID is to be searched for in the entire MTP. If the RefID is found, it refers to the parent element and the CommandExt attribute is addressed from there. Since there can be several CommandExt variables - one for each services, they must be differentiated from each other. Therefore, the name of the individual is extended by the name of the associated service.

Due to limited space, not all mapping rules can be explained in detail here. All rules are available in this project's Github repository\footnotemark{}.
\footnotetext{https://github.com/hsu-aut/mtp2skill}

\begin{figure*}[t]
    \begin{minipage}{\textwidth}
\begin{lstlisting}[
        caption={RML mapping rule to create skills from MTP procedures},
        label={lst:mapping},
        numbers=left,
        xleftmargin=2em,
        framexleftmargin=2em,
        language=XML]
<#OpcUaVariableSkill-hasSkillCommand-CommandExt>
 a rr:TriplesMap;
 rml:logicalSource [
   rml:source "./Manifest_Mischmodul.xml";
    rml:referenceFormulation ql:XPath;
    rml:iterator "/CAEXFile/InstanceHierarchy/InternalElement[@RefBaseSystemUnitPath='MTPServiceSUCLib/Service']/child::InternalElement[@RefBaseSystemUnitPath='MTPServiceSUCLib/ServiceProcedure']" ];

 rr:subjectMap [
   rr:template "http://www.hsu-ifa.de/ontologies/MTPOWL#{@Name}";
   rr:class Cap:OpcUaVariableSkill ];	
  	
 rr:predicateObjectMap [
   rr:predicate Cap:hasSkillCommand;
   rr:objectMap [
   rr:template "http://www.hsu-ifa.de/ontologies/MTPOWL#{//InternalElement/Attribute[Value=current()/parent::InternalElement/Attribute[@Name='RefID']/Value]/parent::InternalElement/Attribute[@Name='CommandExt']/@Name}_{parent::InternalElement/@Name}";
   rr:class Cap:SkillCommand]].

\end{lstlisting}
  \end{minipage}
\end{figure*}

%% file: 05_evaluation.tex
The mapping rules described in Section \ref{sec:mappingApproach} were implemented in a Java application that can be used as a library, a CLI application or as a web service. In this section, the presented approach is evaluated using our implementation on a laboratory scale modular process plant. This plant consists of four modules that provide unit operations such as mixing, distillation, filtering and bottling. Through the use of MTPs, this plant already employs a vendor-neutral control approach using PLCs of \emph{WAGO} and \emph{ABB} that are controlled via a supervisory control system adapted to using MTPs.


In this evaluation, we map MTPs of multiple modules into our capability/skill model. Afterwards, the services of an MTP are executed as skills via our skill-based manufacturing execution system \emph{SkillMEx}\footnotemark{}.
\footnotetext{https://github.com/aljoshakoecher/SkillMEx}

\subsection{Mapping MTPs into an Ontology}
We executed the previously described mapping approach and examined the automatically generated ontology for completeness. In order to do so, so-called \emph{compentency questions} (CQs) were formulated beforehand. CQs can be used to describe the requirements against an ontology and the answers to such questions represent the problems that an ontology is able to solve \cite{GrFo_TheRoleofCompetency_1995}. In total, we used nine CQs to check the completeness of our mapping approach. Two of these nine CQs are:
\begin{enumerate}[label=\textbf{CQ \arabic*:}, align=left, leftmargin=*]
    \item \emph{Which components belong to a certain module \(\mathcal{X}\)?}\\
    This is a rather simple CQ whose answer can be used to evaluate structural completeness, i.e. compare whether or not all components of a PEA were mapped.
    \item \emph{What is the current state of skill \(\mathcal{Y}\)?}\\
    This CQ requires additional information in order to evaluate the current state in a semantic manner. In our mapping approach, we assigned the possible values of \emph{StateCur} to the according states so that this CQ can be sufficiently answered.
\end{enumerate}
For every CQ, a SPARQL query and an expected result were defined to test for a complete transformation.
Integrating all mapped MTPs into one shared ontology allows querying of a whole plant model with SPARQL. This is impossible with MTPs in their AML form as there is no technology that allows querying multiple AML / XML documents at the same time.


\subsection{Executing MTP Services as Skills}
One of the objectives of this publication is to use the services of an MTP together with skills in a uniform and interoperable manner. For evaluation purposes, we integrated our mapping application into \emph{SkillMEx}. Within SkillMEx, users can upload an MTP and execute a mapping. After this mapping has finished, the generated result is automatically added to the existing ontology of SkillMEx.

An MTP skill is displayed with an identical user interface next to previously registered skills. The user interface for invoking state transitions and passing skill parameters or retrieving skill result values is identical for all skill types because it is based on the meta model of skills. It does not differ from skills that are invoked with other technologies (e.g. via web services) or implemented using our previous approaches (see \cite{KHC+_AutomatingtheDevelopmentof_2020} and \cite{KJF_AMethodtoAutomatically_2021}).

In order to invoke MTP services as skills, we implemented an additional \emph{MTP Skill Executor} which is able to invoke transitions and monitor state changes based on integer variables (cf. model extensions in Section~\ref{sec:fundamentals}).
In order to execute any skill, only the skill's IRI, the desired transition as well as the parameters have to be submitted. The MTP Skill Executor queries all other information that are needed to execute the skill such as OPC UA node IDs for all variables as well as the integer values for the desired transition. This information is then used to fire a transition. An OPC UA subscription is created to monitor the variable representing the current state.
With this method, the mixing service of the corresponding module of our laboratory plant could be controlled in a skill-based manner.

%% file: 06_summaryOutlook.tex
This contribution presented a comparison between MTPs and a semantic capability/skill model along with an automatic transformation of MTPs into an existing capability/skill ontology. 
For this transformation, we extended our ontology with a new type of skills and added model elements to describe an MTP's interaction mechanism based on OPC UA variables. Mapping rules using the RDF Mapping Language (RML) were implemtented in a mapping application. Both the ontological model as well as the mapping application are available online.

Skills mapped from an MTP can be used in our skill based manufacturing execution system \emph{SkillMEx} in the same way as other skills even though they use different execution technologies. 
This is possible because all skills share one common interface description which is defined by the ontological skill model.
By integrating MTPs of multiple modules into one ontology, a complete plant can be queried for certain information using the powerful SPARQL query language.

It should be noted that this contribution does not provide a semantic model for the terms and relations of VDI 2658 - instead it provides a transformation into a skill ontology which is built upon different standards and therefore with other terminology. Furthermore, we focused on transforming the services defined in an MTP and additionally transformed data objects (such as sensors and actuators). Additional parts of an MTP such as HMIs or alarms might be transformed in future works to be able to apply advanced features of ontologies such as reasoning or constraint validation.

As various working groups have been standardizing the MTP for a couple of years, it is no surprise that the MTP's level of detail exceeds the one of our capability/skill model. With this contribution, we cannot and do not want to replace MTPs, but instead want to show benefits of more formal modeling approaches and provide an automated solution to bridge the gap between MTPs and capabilities/skills.